\documentstyle[aps,multicol,epsfig]{revtex}
\draft
\def\beginwide{
        \end{multicols} \vspace*{-0.5cm} \noindent
        \rule{3.5in}{.1mm}\rule{.1mm}{5mm} \widetext \medskip }
\def\beginwidetop{
        \end{multicols} \vspace*{-0.5cm} \noindent
        \widetext \medskip }
\def\endwide{
        \hspace*{3.35in}~\rule[-5mm]{.1mm}{5mm}\rule{3.5in}{.1mm}
        \begin{multicols}{2} \vspace*{-1.0cm} \noindent }
\def\endwidebottom{
        \begin{multicols}{2} \vspace*{-1.0cm} \noindent }

\newcommand{\beq}{\begin{equation}}
\newcommand{\eeq}{\end{equation}}
\newcommand{\bdis}{\begin{displaymath}}
\newcommand{\edis}{\end{displaymath}}
\newcommand{\bea}{\begin{eqnarray}}
\newcommand{\eea}{\end{eqnarray}}
\newcommand{\barr}{\begin{array}}
\newcommand{\earr}{\end{array}}
\begin{document}

\title{Energy constrained sandpile models}

\author{Alessandro Chessa$^{(1,2)}$, Enzo Marinari$^{(1,3)}$,
 and Alessandro Vespignani$^{(4)}$}

\address{
1)Dipartimento di Fisica, Universit\`a di Cagliari,
Via Ospedale 72, 09124 Cagliari, Italy\\
2)Istituto Nazionale di Fisica della Materia (INFM), Sezione di 
Cagliari, Italy\\
3)Istituto Nazionale di Fisica Nucleare (INFN), Sezione di Cagliari, 
Italy\\
4)International Center for Theoretical Physics (ICTP),
P.O. Box 586, 34100 Trieste, Italy}

\date{\today}                                                 

\maketitle

\begin{abstract}  
We study two driven dynamical systems with conserved energy. 
The two automata contain the basic dynamical rules of the Bak, Tang and 
Wiesenfeld sandpile model. In addition a global constraint on the energy 
contained in the lattice is imposed. In the limit of an infinitely slow 
driving of the system, the conserved energy $E$ becomes the only parameter
governing the dynamical behavior of the system. Both models show scale free
behavior at  a critical value $E_c$ of the fixed energy.  
The scaling with respect to the relevant scaling field points out that the 
developing of critical correlations is in a different universality class 
than self-organized critical sandpiles. 
Despite this difference, the activity (avalanche) probability  
distributions appear to coincide with
the one of the standard self-organized critical sandpile. 
\end{abstract}

\pacs{PACS numbers: 64.60.Lx, 05.40.+j, 05.70.Ln}

%\begin{multicols}{2}
In the study of non-equilibrium critical phenomena, cellular automata
(CA) showing self-organized criticality (SOC) have acquired a very special
role\cite{btw,grin}.  Differently from usual continuous phase
transitions, they would spontaneously evolve into a critical state
without explicit fine tuning of control parameters.  Another reason of
interest lies in the fact that numerical computations based on SOC
ideas have shown that slowly driven systems can lead to a stationary
state with a dynamical activity characterized by avalanches of widely
distributed amplitude\cite{grin}.  Avalanche behavior is a basic
feature of many experimentally observed phenomena ranging from
magnetic systems\cite{bar} to micro-fracturing process \cite{ae} and
earthquakes\cite{earth}.  The prototypical model for SOC is represented
by Bak, Tang and Wiesenfeld (BTW) sandpile automata\cite{btw}, 
in which an infinitesimally slow external
driving of sand particles associated with a threshold rearrangement
dynamics lead to a stationary state with activity (avalanches)
distributed on all length scales\cite{btw}.  More widely, the model is
generalized by identifying the sand grain as energy, stress or
pressure quanta.  In this way the analogy with other physical
phenomena appears more clearly.

Despite the vast activity in the field, the general picture of SOC 
phenomena contains many ambiguities. It has been pointed out by 
several authors\cite{grin,sorn,vz} that the driving rate acts exactly 
as a control parameter that has to be fine tuned to zero in order to 
observe criticality.  For instance, in sandpile the stationary state 
results from the balance of the driving field and dissipation rates 
intrinsically operating because of the system open boundary. 
The critical point is reached only through a limit process in which 
both driving and dissipation rates tend to zero.  This point 
corresponds to a locality breaking of the dynamical rules\cite{vz} 
that determines the onset of the critical correlation 
properties\cite{FOOT-A}.  In 
this framework many relations with non-equilibrium critical phenomena, 
such as adsorbing critical point\cite{dick}, have been enlightened.  
However, many important issues are still open.  It is not 
clear the interplay among the self-organization into a stationary state 
due to the energy 
balance and the dynamical developing of correlations.  
Also numerically many important features, such as critical  exponents, 
universality classes and the upper critical dimension
are very difficult to obtain to a sufficient degree of 
precision\cite{ben,lubeck}.  
This is mainly due to the inherent strong effect of finite 
size corrections present in the original model; the 
boundary size plays an active role, being the only dissipative 
ingredient leading to the stationary state\cite{vz}.

In this letter we introduce a stochastic CA which contains the basic
elements of the sandpile model, but is defined on a lattice 
with periodic boundary conditions,
and has a global constraint in the energy accumulated.  
The average energy contained in
the system is therefore constant and fixed from the outside.  This
resembles a microcanonical definition of the sandpile automata.  The
reason for studying this model is two-fold.  First it seems more
appropriate to represent some phenomena in which the 
dissipation is not intrinsically linked to the activity of 
the systems.  The second is
that it could shed light on many properties of the SOC sandpile by
allowing its analysis in a framework which look closer to usual
statistical physics.  Finally, it turns out that SOC and
microcanonical sandpile appear to share the same critical exponents for the
avalanche distribution.  The latter does not suffer heavily of
finite-size correction effects because of the possibility of using
effectively periodic boundary conditions.  Thus, the microcanonical
sandpile could be used to settle universality
classes and upper critical dimension issues.

We consider two models with conserved energy.  In both we start from a 
given sand configuration $\{ e_i \}$, that can be random or 
the result of a former run (if needed after modifying its energy), where 
$i=(x,y)$ labels the $L^2$ sites of a $2-d$ lattice with periodic boundary 
conditions.  The total amount of sand (the energy of the system) is 
$E\equiv \sum_i e_i$.  The system is supposed to be in contact with an 
external reservoir with which it can exchange energy in both 
directions; an incoming as well as an outcoming energy flux is 
present.  We think at the system in a sort of thermal equilibrium with 
a fixed value of energy.  This implies the two fluxes on average are 
equal.

In both models, the first stage of the dynamics is the
addition of an energy unit on a random site. In order to preserve
the total amount of energy, we have to introduce an extraction flux 
that takes away one unit of energy from the system. 
We do that in two ways. In the
first model (that we call {\em Random Subtraction}, RS) we take away
one unit of energy in a random site: this model is discrete, and $e_i$
can only take integer values from $0$ to $4$. In the second model
(with a {\em Continuous Subtraction}, CS) we have an homogeneous
dissipation, where each site looses energy proportionally to the local
energy density. Here we transform $e_i \to e_i \frac{E}{E+1}$.  The
basic variables of this second model are continuous, and can take
values between $0$ and $4$.

The internal dynamics of both models is supposed to be very fast with 
respect to the energy addition and extraction rates, in analogy with 
the slow driving assumption commonly used in SOC models.  After 
the energy addition and extraction,the avalanching process follows in 
the usual way.  If $e_i$ is larger or equal to $4$ (the critical 
threshold for local stability), the energy on the site gets shared 
among the nearest neighbors sites, and the avalanche is followed since 
a stable state is reached.  After the avalanche stops, the update 
continues with the deposition and extraction of a new energy unit.

We impose periodic boundary condition on the system, i.e., 
$e(i,0)\equiv e(i,L+1)$ and $e(0,j)\equiv e(L+1,j)$ . In usual sandpile 
this would lead to troubles because $E$ can only increase. Thus after 
a finite time a state with an infinite avalanche that goes on forever 
is reached. This state obviously is not related with the critical 
stationary one. For this reason periodic boundary conditions have  
never been used to determine the critical properties of sandpile 
models. The price to pay for that is the inclusion of the strong 
finite size corrections induced by the finite boundary dissipation.

In these models the energy dissipation is acting as an independent 
driving, while in usual sandpile is always linked to the toppling 
event itself. In SOC sandpile also  the average energy is dependent upon 
driving and dissipation because of the energy balance while in our 
microcanonical model this self-organization is ruled out.
Thus in these models, the total energy $E$ is a free
parameter, that can be freely tuned.
Here we will mainly  present  the CS model and some evidences for an
analogous behavior of the RS discrete model, where 
the critical energy density turns out to
coincide with the stationary energy density of the BTW model.

We study the CS sandpile  model 
as a function of the control parameter $E$: we start
with small $E$ (few energy) and small correlation length,
and we increase $E$  keeping the correlation length smaller that the 
lattice size in order to make finite size effects small 
(we present here only results
that do not change when going from $L=256$ to $L=512$). When $E\to
E_c$ the average avalanche size $\langle s \rangle$ diverges, together
with the average avalanche duration $\langle t \rangle$, its gyration
radius $\langle r \rangle$ and the average number of {\em different}
sites touched during an avalanche, $\langle s_d \rangle$.  The system
reaches in this case a critical point, that we will show to be
characterized by BTW exponents.
We have determined numerically the probability
distributions $P_s(s)$, $P_t(t)$ and $P_{s_d}(s_d)$, and determined
the exponents of their asymptotic power law decay.  

\begin{figure}[htb]
\centerline{\hbox{
\epsfig{figure=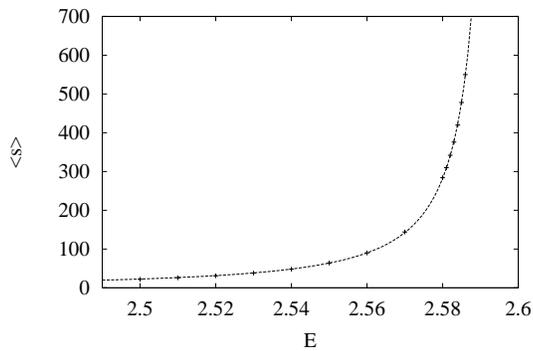,width=5cm,angle=-90}
}}
\caption{$\langle s \rangle$ versus $E$, with the best fit to a power
    divergence.}
\label{F-S}
\end{figure}

In fig. (\ref{F-S}) we show the avalanche average size, $\langle s
\rangle$, for $L=512$, as a function of $E$, together with the best
fit to a simple power divergence (done by using all the points plotted
in the figure). We fit the asymptotic behavior:

\begin{equation}
  \langle s \rangle\sim \frac{1}{(E_c-E)^\gamma}\ ,
\end{equation}
and we find $E_c=2.596 \pm 0.001$ and $\gamma=1.41\pm 
0.03$\cite{errors}. 
The average avalanche size can be shown to scale asymptotically 
as the system response function $\chi_E$,
that implies  $\chi_E\sim (E_c-E)^{-\gamma}$.
The latter expression characterizes how the system
reacts to external perturbations\cite{vz}.

\begin{figure}
\centerline{\hbox{
\epsfig{figure=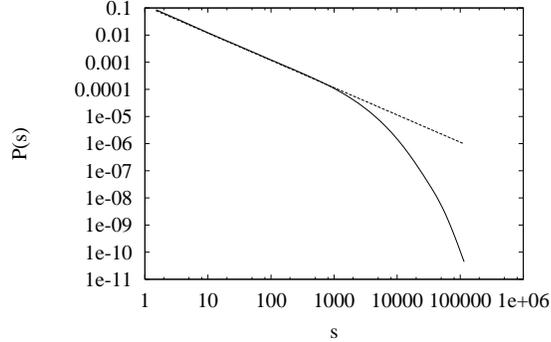,width=5cm,angle=-90}
}}
\caption{$P_s(s)$ versus $s$ in log-log scale (the solid line is a smooth
  interpolation of the numerical data) and best power fit (see text).}
\label{F-PS}
\end{figure}

For the energy range where $\langle s \rangle>20$ we have computed an
effective, energy dependent power exponent for the avalanche 
distribution. We show in figure
(\ref{F-PS}) the typical situation (for $L=512$, at $E=2.586$): since
we are not at $E_c$ the power law decay is truncated (at a value that
turns out to be of order $\langle s \rangle$). We always fit the power
law, $P_s(s)\sim s^{-\tau_s}$ in a range of $s$ that goes from $1$
to $\langle s \rangle^{(L)}$. One sees from the figure that the fit
(the dashed straight line) is very good on three decades (the solid
line is a smooth interpolation to the numerical data).

The exponents one finds at finite $(E_c-E)$ have to be extrapolated to
the critical point.  We fit $\tau_s$ to an asymptotic value with
corrections linear in $\log(E_c-E)^{-1}$ (by following Manna
\cite{manna1}): we find $\tau_s(E_c)=1.26\pm 0.02$, where again the
error is only statistical.
Still, in the limit of such a statistical accuracy (that is of the
same level it can be reached for the BTW model), we find the
same exponent that is believed to describe the BTW scaling.
The same procedure works for the time duration of an avalanche. Here
by assuming that
\begin{equation}
  \langle t \rangle\sim \frac{1}{(E_c-E)^\theta}\ ,
\end{equation}
we find a very good best fit with $E_c=2.597\pm 0.001$ and
$\theta=0.80\pm 0.04$. With the same approach used for $P_s(s)$ we find
that $P_t(t)\sim t^{-\tau_t}$, where $\tau_t(E_c)=1.49\pm
0.04$. It is worth to remark that in measuring the time duration of an 
avalanche  different definitions of time can be used. Here we adopt 
the one commonly implemented in SOC automata: at each integer time-step
all currently active sites topples.
Again, in the error bars given by the fitting procedure, we find a
remarkable agreement with the $\tau_t=\frac{3}{2}$ that 
one expects for the BTW model. 
The same procedure applied to the different sites touched from
an avalanche, $\langle s_d \rangle$, leads again to a divergence at
$E_c=2.597$ with an exponent equal to $1.34\pm 0.03$. Again,
$P_{s_d}(s_d)$ shows a clear power law behavior, and we find that
$\tau_{s_d}(E_c)=1.27\pm 0.04$, again in good agreement with the
BTW result.

\begin{figure}
\centerline{\hbox{
\epsfig{figure=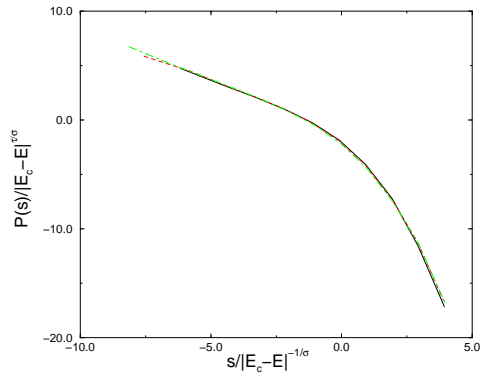,width=5cm,angle=-90}
}}
\caption{Scaling plot of $P(s)/(E_{c}-E)^{\tau/\sigma}$ 
versus $s/|E_{c}-E|^{-1/\sigma}$.}
\protect\label{F-DS}
\end{figure}

We can also define a natural characteristic length in the system. In 
general, close to the critical point the avalanche distribution has 
the scaling form
 
\begin{equation}
   P(s)= s^{{-\tau}}G(\frac{s}{s_{c}})\ ,
\end{equation}
where $G(x)$ is a universal function and $s_{c}$ is the avalanche
cut-off size.  The latter is the system characteristic length that
close to the critical point scales as $s_{c}\sim (E_{c} -
E)^{-1/\sigma}$.  In order to test the scaling assumption and find an
estimate of the $\sigma$ exponent we have used a data collapse
technique.  For energy values close to the critical one, the plot of
$P(s)/(E_{c}-E)^{\tau/\sigma}$ as a function of the rescaled variable $s/
(E_{c}-E)^{-1/\sigma}$ must collapse into the same universal curve by
using the correct values of $\tau$ and $\sigma$.  In fig.
(\ref{F-DS}) we show the data collapse from avalanche distributions
obtained with $(E_c-E)$ ranging over almost one order of magnitude.
The values we obtain for the exponents are $\tau=1.20\pm 0.05$ and
$\sigma=0.55\pm 0.03$.  By using Eq.(3), we can immediately write the
relation:

\begin{equation}
<s>=\int s^{-\tau+1}P(s) ds \sim (E_c-E)^{\frac{\tau-2}{\sigma}} ,
\end{equation}
which immediately gives the scaling relation $\gamma=(2-\tau)/\sigma$.
The latter relation is satisfied by the exponent values we obtain,
providing a further consistency check for the numerical results.  It
is interesting to remark that the values of the critical exponent
$\gamma$ and $\sigma$ are very different from those obtained in the
SOC sandpile ($\gamma=1$, $\sigma=0.77$ \cite{CMVZ}).  In the SOC
sandpile model these exponents are defined with respect to the
dissipation rate which plays the role of the control
parameter\cite{vz}.

>From the previous analysis we can therefore identify two main
dynamical mechanism in SOC models.  The first is the self-organization
that is driven by the energy balance condition. The sandpile evolves
in order to set its energy density so that the avalanche finds a
background that allows it to dissipate enough energy. This process
does not imply criticality. The second mechanism is the dynamical
interaction which builds up in the system the long range correlations
which on its turn create the critical avalanche distribution. This is
just in the presence of the locality breaking obtained in the limit of
infinite slow driving of the system.  In our microcanonical version of
the model, we control the energy self-organization from outside. The
critical point is thus reached just in the presence of the critical
energy density which allows the slow driving to generate the critical
configuration for the system. These two different ways of reaching the
critical point appear to generate different scaling properties with
respect to the control parameter. On the contrary the scaling behavior
properties right at criticality result to be the same, within the
numerical accuracy of our simulations, in both SOC and energy
constrained sandpiles automata. This suggest the two systems build up
differently critical correlations, eventually leading to a critical
point which present the same avalanche distribution.

Finally we report that the RS model shows the same kind of
behavior. It is crucial to note that here we find $E_c=2.127\pm 0.004$
to be compared with the $E_c=2.125$ that Grassberger and Manna
\cite{grasma} find for BTW. The energy where the model becomes
critical is exactly the energy reached from BTW in the steady state.
This is because the RS model has a microscopic dynamics which is
identical to the SOC BTW model. The difference is in the the way the
system is driven to criticality and thus in the energy constraint.
This allows us to compare directly quantities right at the critical
point, that should assume the same values in both models.  The
critical behavior of this model is characterized by the same critical
exponents of the CS model. The detailed presentation of these
numerical data will appear in a forthcoming paper\cite{CMVII}.

We are grateful to R. Dickman and S. Zapperi for useful discussions and
a careful reading of the manuscript. The main part of the numerical
simulations have been run on the {\em Kalix} parallel computer
\cite{KALIX} (a Beowulf project at Cagliari Physics Department). We
thank Gianni Mula for leading the effort toward organizing this
computer facility.

%\end{multicols}

\begin{thebibliography}{99}
\bibitem{FOOT-A}
Sandpile models, for instance, are driven by adding a single energy 
grain on a randomly chosen site, when no active sites are present.  In 
this way, avalanches are instantaneous with respect to the driving 
time scale.  Non-locality is thus implicitly enforced in computer 
simulations, where the evolution of a single site depends on the state 
of the entire system.
\bibitem{errors}
The error we report is the  statistical one obtained from the fitting 
procedure. The true uncertainty on the result is also affected by 
the systematic error due to the deviation from scaling, 
which, however, is difficult to estimate.
\end{thebibliography}
\end{document}